# Radio and infrared emission from a [WC]-type planetary nebula in the LMC

Albert A. Zijlstra[1], Peter A.M. van Hoof[2], Jessica M. Chapman[3] and Cecile Loup[4]

[1] European Southern Observatory, Karl-Schwarzschild-Strasse 2, D-85748 Garching bei München, Germany
[2] Kapteyn Astronomical Institute, P.O. Box 800, NL-9700 AV Groningen, the Netherlands
[3] Australia Telescope National Facility, P.O. Box 76, Epping, NSW 2121, Australia
[4] Space Research Groningen, P.O. Box 800, NL-9700 AV Groningen, the Netherlands

**Summary.** Radio continuum emission has been detected from a planetary nebula in the Large Magellanic Cloud: this is the first radio continuum detection for any planetary nebula outside our galaxy. The radio flux density is a factor of two lower than predicted from the H$\beta$ flux. This could be due either to a two-component morphology or to a stellar contribution to the H$\beta$ emission. We have modeled the optical and infrared spectrum using the photo-ionisation code Cloudy: a very good fit is obtained if we assume the nebula is carbon rich. The derived diameter implies an evolutionary age of the nebula of $< 1000\,\rm yr$, similar to what is derived from the IRAS colours alone. The central star, which is of [WC] type, has a much higher stellar temperature than Galactic [WC] stars showing similar circumstellar IRAS colours. An explanation could be that the expansion velocity of the nebula is lower than those of its galactic counterparts. This radio detection indicates that accurate nebular luminosities could be determined from their radio emission for many LMC planetary nebulae.

## 1. Introduction

Radio and infrared observations have become important tools in the study of galactic planetary nebulae (PN). Radio continuum emission has the advantage over optical emission that it is not affected by atmospheric or interstellar extinction. The radio continuum emission can be used to detect PN in obscured Galactic regions (e.g. van de Steene and Pottasch 1993) and to determine accurate nebular luminosities. The infrared (IRAS) region contains the dust emission which can contribute up to 80 per cent of the total PN energy budget. From the IRAS colours the dust temperature can be derived, giving an indication of the nebula's physical size.

The study of galactic PN is severely hampered by the lack of accurate distances. Only the bulge provides a sample at a well-known distance, but it contains only low-mass planetary nebulae with old progenitors (e.g., Pottasch and Acker 1989). More luminous, high-mass PN with young progenitors are not found in the bulge and, for such objects, the Magellanic clouds have become the testing ground.

Radio observations of LMC PN have not yet been attempted due to the need for a sufficiently sensitive instrument. IRAS data have been available (Schwering 1989) but have not been used, mainly because until recently the optical positions for LMC PN were too inaccurate to establish reliable associations. Also, the flux densities are at the limit of IRAS detectability and require special processing. As a result, the study of LMC PN has exclusively been based on optical and, in a few cases, IUE observations. The optical measurements are not without problems: the observations must be done at significant air masses ($> 1.2$), requiring a correction for atmospheric dispersion. In addition, the fields are generally very crowded. The difficulty of obtaining reliable reddening corrections adds to the uncertainty: recent observations indicate reddening is significant for some objects (e.g., Vassiliadis et al. 1992a).

In this paper we present the first radio detection of a PN in the LMC. This PN (SMP58) was also detected by IRAS. We discuss its characteristics and compare it to other IRAS-detected PN in the LMC. SMP58 contains a Wolf-Rayet-type central star (one of four such objects known in the LMC), which makes it especially interesting. These low-mass stars show spectra similar to those of the massive WR stars, with strong carbon and helium emission lines. Following van der Hucht et al. (1981), they are denoted as [WC] stars, to distinguish them from their massive counterparts. [WC] stars occur in approximately 5 per cent of all galactic PN (taken from Acker et al. 1992); the four LMC [WC] stars constitute a similar percentage of spectroscopically observed PN in the LMC. They are generally assumed to be helium-burners with little or no hydrogen envelope: at present there are no theoretical model calculations available for the evolution of such stars. We show that SMP58 has bluer IRAS colours than galactic [WC] stars of similar temperature, indicating a more compact nebula. This conclusion is confirmed by modeling of the spectrum, and possibly indicates lower nebular expansion velocities.

## 2. Observations

We observed the LMC planetary nebula SMP58 with the compact array of the Australia Telescope on two occasions. The array consists of six 22-m antennas: five are movable on a 3 km east–west baseline and the sixth antenna has a fixed position 3 km further west. During the first observations, the five antennas were placed in a highly compact configuration. The sixth antenna was not used in the final analysis because of the uneven $uv$ coverage it would create. There were problems with the amplitude calibration and we reobserved the source in June, using a more extended configuration which gave higher resolution and allowed use of all antennas. All observations were done simultaneously at 3 and 6 cm, in blocks at several different hour angles to improve the $uv$ coverage. The pointing was offset from the PN to avoid problems with DC offsets which can cause artifacts at the field centre. The observations are summarised in Table 1.

Data reduction was done using the AIPS software package. The central 75% of the band was averaged in frequency to obtain a continuum database. A large map was made in order to find confusing sources, which were subsequently removed through the CLEAN procedure. The flux was determined as the average between the peak in the cleaned map, and the value obtained from a gaussian fit to the source. The flux determinations of the June data are expected to be accurate to better than 10 per cent, in addition to the uncertainty introduced by the the rms noise of the data. The uncertainty is primarily due to the poor $(u, v)$ coverage and the low signal to noise of the detections. In addition, the flux density of the primary calibrator is uncertain by a few per cent: this introduces a systematic uncertainty in the derived flux densities.

**Table 1.** AT observations of SMP58

|  | May 1993 |  | June 1993 |  |
| --- | --- | --- | --- | --- |
| AT configuration | 0.375 |  | 6A |  |
| wavelength (cm) | 3 | 6 | 3 | 6 |
| frequency (MHz) | 8640 | 4800 | 8640 | 4800 |
| bandwidth (MHz) | 128 | 128 | 128 | 128 |
| $T_{\rm sys}$ (K) | 50 | 50 | 50 | 50 |
| flux calibrator | 1934−638 | 1934−638 | 1934−638 | 1934−638 |
| calibrator flux (Jy) | 2.59 | 6.34 | 2.59 | 6.34 |
| phase calibrator | 0454−810 | 0454−810 | 0522−611 | 0522−611 |
| int. time (min) | 90 | 90 | 100 | 100 |
| resolution ($''$) | 35 × 12 | 72 × 22 | 2.2 × 1.6 | 5.3 × 3.7 |
| rms noise (mJy/beam) | 0.16 | 0.20 | 0.10 | 0.10 |
| PN flux (mJy) | 0.6: | 1.1: | 0.84 | 0.79 |
| uncertainty (mJy) |  |  | 0.13 | 0.13 |
| radio position (J2000) | $\alpha = 05^{\rm h}24^{\rm m}20.7^{\rm s}$ |  | $\delta = -70°05'01.5''$ |  |

## 3. The radio–H$\beta$ discrepancy

We detected radio emission from SMP58 at a level of about 0.8 mJy at 3 and 6 cm. The 3 cm image is shown in Fig. 1. The optical position (Meatheringham and Dopita 1991a) of $\alpha = 5^{\rm h}24^{\rm m}20^{\rm s}$, $\delta = -70°05'03''$ (J2000) agrees to within $2''$ with the more accurate radio position, leaving no doubt that the detected radio emission originates from the planetary nebula. The fact that the radio flux densities at 3 and 6 cm are identical within the uncertainties indicates that the radio emission is optically thin at these wavelengths. In this case the radio flux density and the H$\beta$ flux should be related by a simple relation (Pottasch 1984, eq. IV-26) that only requires knowledge of the electron temperature and the amount of singly and doubly ionised helium; these parameters are given by Monk et al. (1988), Meatheringham et al. (1988) and Meatheringham and Dopita (1991a): $T_{\rm e} = 11200$ K, $n({\rm He^+})/n({\rm He^0}) = 0.072$, $n({\rm He^{++}}) = 0$. The H$\beta$ flux predicted from this relation and our radio flux is $\log(F_{\rm H\beta}) = -12.63 \,{\rm erg\,cm^{-2}\,s^{-1}}$. This is considerably smaller than the observed value of $\log(F_{\rm H\beta}) = -12.20$, where we have corrected for the reddening: $c = 0.28$, derived from the H$\alpha$/H$\beta$ ratio (Meatheringham and Dopita 1991a). Conversely, the observed H$\beta$ flux with the H$\alpha$/H$\beta$ reddening would predict a radio flux density of 1.9 mJy.

The reason for this large discrepancy is not clear. Obviously both the radio and optical measurements should be repeated: if the 3 cm flux density were found to be be less than 0.5 mJy, it would change our conclusion that the nebula is optically thin. This error is larger than the estimated uncertainty on the flux, but the possibility makes additional observations desirable. A smaller optical extinction would also lessen, but not remove, the discrepancy.

A possible explanation for the discrepancy could be that the nebula has a very compact core which is highly optically thick to radio continuum emission. A Galactic example of such an object is the planetary nebula M2-9 where the core is so optically thick that it contributes very little flux at 5 GHz: most of the detected radio flux comes from more extended emission (Zijlstra et al. 1989). The optically thick component only becomes apparent at higher frequencies. Our present results do not allow us to confirm such a core–halo morphology for SMP58. An alternative explanation could be that the observed H$\beta$ flux is contaminated by line emission from the central star. This can be tested through measurements of other Balmer lines, which in such a case would yield different values for the reddening.

The observed flux density can be used to obtain limits to the size of the nebula. The fact that the radio brightness temperature cannot exceed the electron temperature allows us to derive a lower limit to the diameter of the full nebula of $0.08''$ (e.g. Kwok 1986). If an optically thick core is present we can similarly put limits on the core size, using the fact that for optically thick emission the brightness temperature is equal to the electron temperature. Assuming that such a core contributes less than $\sim 0.1$ mJy at 6 cm, we derive an upper limit to its diameter of $0.03''$. Using optical measurements, Dopita and Meatheringham (1991) model the nebula as a very thin shell with an inner radius of 0.019 pc and an outer radius of 0.022 pc. The corresponding predicted angular diameter is $0.15''$. This diameter, though possibly not the density structure, is in agreement with our limit on the size of the full nebula of $> 0.08''$.

In the following section we will discuss IRAS data on PN in the LMC. Using the IRAS data for SMP58, we will then model the optical and infrared spectrum to show that the nebula is possibly more compact than found by Dopita and Meathering-ham.

**Table 2.** IRAS sources associated with LMC pn

|        | $F(12\mu)$ (Jy) | $F(25\mu)$ (Jy) | $F(60\mu)$ (Jy) | $\log F(H\beta)^a$ | $c$   | $\log(N_e)$ (cm$^{-3}$) | ref    | $\log L_\star(L_\odot)^b$ | IRE  |
|--------|-----------------|-----------------|-----------------|--------------------|-------|-------------------------|--------|---------------------------|------|
| SMP1   | 0.06            | 0.24            | 0.14            | $-12.46$           | 0.17  | 3.56                    | 1,2    | 3.8                       | 1.3  |
| SMP6   | <0.06           | 0.11            | <0.16           | $-12.67$           | 0.50  | 4.25                    | 1,2,3,7| 3.9                       | 0.3  |
| SMP8   | 0.21            | 0.50            | –               | $-13.74$           | 0.23  | 3.74                    | 1,2,10 | 2.5                       | 56   |
| SMP11  | 0.18            | 0.30            | –               | $-13.15$           | –     | –                       | 5      | –                         | –    |
| SMP28  | <0.24           | 0.44            | –               | $-13.35$           | –     | –                       | 5      | –                         | –    |
| SMP31  | 0.13            | 0.45            | <0.12           | $-12.91$           | <0.88 | 4.0                     | 1,2,9  | <4.0                      | >1.3 |
| SMP58  | 0.15            | 0.22            | –               | $-12.48$           | 0.28  | 4.82                    | 1,2    | 3.9                       | 1.5  |
| MG45   | 0.15            | 0.32            | 0.16            | –                  | 0.90  | –                       | 2,6    | –                         | –    |
| SMP61  | 0.08            | 0.13            | <0.16           | $-12.48$           | 0.20  | 4.52                    | 1,2,3  | 3.8                       | 1.0  |
| SMP85  | 0.10            | 0.38            | <0.75           | $-12.42$           | 0.25  | >5.0                    | 2,4    | 3.9                       | 1.4  |
| SMP98  | 0.16            | 0.23            | 0.14            | $-12.52$           | 0.39  | 3.86                    | 4      | 3.9                       | 1.3  |
| MA18   | 0.08            | 0.12            | <0.14           | $-12.67$           | 0.07  | >5.0                    | 4,8    | 3.4                       | 2.0  |

$^a$ H$\beta$ fluxes are not corrected for reddening.
$^b$ The stellar luminosities are calculated as $150 \times L_{H\beta}$.
$^c$ The IRE is derived by simply summing over the IRAS bands and should therefore considered as lower limits, especially for young objects.
references: 1. Meatheringham et al. 1988; 2. Meatheringham and Dopita, 1991a; 3. Meatheringham and Dopita, 1991b; 4. Vassiliadis et al. 1992a; 5. Wood et al. 1987; 6. Morgan and Good (1992); 7. The position in Meatheringham and Dopita (1991b) is wrong. The correct position is given in Meatheringham and Dopita (1991a); 8. Vassiliadis et al. 1992b; 9. Extinction from Monk et al. (1988) is considered as upper limit; 10. The H$\beta$ flux may be in error.

## 4. IRAS detections of LMC planetary nebulae

The IRAS detection of SMP58 is given in Schwering's catalogue (1989, here after S89) of IRAS sources in the LMC, with flux densities of 0.15 and 0.22 Jy at 12 and 25$\mu$m. The IRAS and radio positions agree within 11″, leaving little doubt about their association. The flux densities are reasonable: the infrared excess (IRE, defined as the ratio of the infrared flux over the Ly$\alpha$ flux) is about 1.4. Here we determine the IRE by simply summing the emission in the IRAS bands. Strictly speaking this should be interpreted as a lower limit, since (especially for young objects) significant dust emission arises shortward of the 12$\mu$m band. We use this value, however, because it can be compared to determinations for Galactic PN. In the Galaxy the IRE shows a radial gradient coincident with the Galactic metallicity gradient. The value of 1.4 coincides with the average for Galactic PN outside the solar circle and is about a factor of two below PN in the inner disk (Zijlstra 1990).

S89 lists in total 29 LMC PN with possible IRAS associations. This listing can now be significantly improved because of (1) the much more accurate positions for PN now available, and (2) the availability of the Faint Source Catalogue (FSC) which covers the full LMC (S89 only covers the inner regions). We have correlated both the FSC and S89 catalogues with the list of LMC PN and selected all sources where the IRAS and PN position agree to within 15″: in total 12 associations were found, with IRAS flux levels between 0.1 and 0.5 Jy. These sources are listed in Table 2. Several of the associated IRAS sources appear in both the FSC and S89. In such cases the two catalogues show generally good agreement in flux densities and positions; the flux densities given in Table 2 are those from the FSC. Of the 12 sources in Table 2, only SMP58 is not listed in the FSC. The large fraction of sources from the FSC is possibly caused by the fact that in confused regions, where the S89 catalogue is better, the positional accuracy is lower.

The IRAS colours indicate that the IRAS-detected nebulae are very young objects. The $F_{12}/F_{25}$ ratios are in the range 0.25–0.7 and are significantly bluer than the majority of Galactic PN (Pottasch et al. 1988). This indicates that we have selected objects which have evolved rapidly towards the PN phase: the $F_{12}/F_{25}$ ratio drops rapidly during the early evolution after the end of the AGB mass loss as a result of the decreasing dust temperature. At later phases the peak of the dust emission occurs longward of 25$\mu$m, and the $F_{12}/F_{25}$ ratio is less affected by further evolution; most of the Galactic PN are in this phase. The blue IRAS colours in the LMC are the results of a selection effect, in the sense that a 12$\mu$m detection is in general necessary in order to get a sufficiently accurate IRAS position, combined with the fact that at the distance of the LMC the peak of the PN luminosity function is not far above the detection limit of IRAS. However, it is interesting that such rapidly evolving objects exist at all in the LMC. The short post-AGB time scales are consistent with the high electron densities of most of the objects.

The objects SMP58 and SMP61 have [WC] central stars (Monk et al. 1988, Meatheringham and Dopita 1991a), based on the presence of the CIV $\lambda$5806 line. Meatheringham and Dopita also suggest that SMP31 may have a [WC] central star, because the individual Balmer lines give inconsistent values for the extinction. Interestingly, these objects are among the high-luminosity PN in Table 2. As only four [WC] stars are known in the LMC (Monk et al. 1988), not including SMP31, it appears that an unexpectedly high fraction of those have been detected by IRAS. It is noteworthy that one non-detected source (SMP38) has a somewhat lower nebular density than the detected two, while the other (LM1-64, Peña et al. 1994), is probably more evolved because of its higher stellar temperature.

## 5. A model for the optical, IRAS and radio data of SMP58

An optical spectrum of SMP58 is given by Meatheringham and Dopita (1991a), We have attempted to model this spectrum using a modified version of the photoionization code Cloudy 84.06 (Ferland 1993). The following parameters were kept free in the modeling: the stellar temperature and luminosity, the inner and outer radius of the shell, the density, chemical composition and dust-to-gas ratio of the shell and the distance to the nebula. The distance is used as an independent check on the model, since the distance to the LMC is very well known. A detailed discussion of the method with special emphasis on the distance determination will be given in a forthcoming paper.

The shell is assumed to be spherical and to have uniform density. For the star we assumed a blackbody spectrum. As input to the model we use line ratios from the optical spectrum, the IRAS flux densities and the radio fluxes. The HeII $\lambda4686$ line could not be reproduced by the fit. It is probably contaminated by line emission from the central star, since [WC] stars are expected to show this line in emission. Possible contamination of the HI, HeI and CII lines was not considered. We also ignored possible NIII contamination of the H$\delta$ line.

**Table 3a.** Observed and predicted quantites for SMP58

| $\lambda$ (Å) | id | observed | Cloudy | Siebenmorgen |
|---|---|---|---|---|
| 3727 | [O II] | 8.2 | 9.6 | |
| 3869 | [Ne III] | 33.9 | 34.2 | |
| 4102 | H$\delta$ | 25.8 | 29.8 | |
| 4267 | C II | 1.4 | 1.4 | |
| 4340 | H$\gamma$ | 47.2 | 50.6 | |
| 4363 | [O III] | 9.0 | 11.6 | |
| 4472 | He I | 6.2 | 4.2 | |
| 4686 | He II | 1.8 | 0.6 | |
| 4861 | H$\beta$ | 100.0 | 100.0 | |
| 4959 | [O III] | 244.7 | 247.9 | |
| 5007 | [O III] | saturated | 743.7 | |
| 5876 | HeI | 13.1 | 14.7 | |
| 6300 | [O I] | 2.4 | 1.6 | |
| 6312 | [S III] | 1.9 | 1.9 | |
| 6364 | [O I] | 1.0 | 0.5 | |
| 6548 | [N II] | 2.0 | 2.0 | |
| 6563 | H$\alpha$ | saturated | 272.3 | |
| 6583 | [N II] | 5.8 | 5.9 | |
| 7065 | He I | 11.1 | 11.2 | |
| 7136 | [Ar III] | 7.7 | 7.7 | |
| 7320 | [O II] | 6.5 | 7.5 | |
| 7330 | [O II] | 6.8 | 6.1 | |
| 12 $\mu$m | IRAS | 0.15 Jy | 0.10 Jy | 0.13 Jy |
| 25 $\mu$m | IRAS | 0.22 Jy | 0.29 Jy | 0.24 Jy |
| 60 $\mu$m | IRAS | — | 0.12 Jy | 0.12 Jy |
| 3 cm | radio | 0.84 mJy | 0.86 mJy | |
| 6 cm | radio | 0.79 mJy | 0.84 mJy | |

**Table 3b.** model parameters for SMP58

| parameter | value | element | abundance |
|---|---|---|---|
| $T_*$ | 57300 K | $\epsilon$(He) | 10.94 |
| $L_*$ | 5330 L$_\odot$ | $\epsilon$(C) | 9.22: |
| $r_{\rm in}$ | 0.0059 pc | $\epsilon$(N) | 7.14 |
| $r_{\rm out}$ | 0.0237 pc | $\epsilon$(O) | 8.29 |
| $\log(n_{\rm H})$ | 4.95 [cm$^{-3}$] | $\epsilon$(Ne) | 7.12: |
| distance | 50.7 kpc | $\epsilon$(S) | 6.39 |
| | | $\epsilon$(Ar) | 5.78 |
| $T_{\rm e}$ | 12000 K | | |
| $M_{\rm sh}$ | 0.164: M$_\odot$ | | |
| $r_{\rm ion}$ | 0.0101 pc | | |
| $\Theta_{rad}$ | 0.041″ | | |
| $M_{\rm ion}$ | 0.011 M$_\odot$ | | |
| IRE[a] | 5.5 | | |
| log F(H$\beta$) | $-12.54$ erg cm$^{-2}$s$^{-1}$ | | |

[a] Here the IRE is calculated as the integral over the dust emission spectrum; the value can therefore not be directly compared with those in Table 2.

A comparison of the observed and predicted quantities can be found in Table 3a. The resulting parameters for the optimum fit are given in the Table 3b. The abundances are logarithmic number densities with respect to hydrogen ($\epsilon$(H) $\equiv$ 12.00). As additional information we also give the mean electron temperature, the total shell mass, the ionisation radius, the angular radius, the ionised mass, the infrared excess and the (non-reddened) H$\beta$ flux.

The observed line ratios and the radio flux densities are well reproduced by the fit. Also, the predicted distance is consistent with the known distance to the LMC, which is an important validation of the model. The outer radius is determined exclusively from the IRAS fluxes and is therefore not well determined. However, the model does not reproduce the IRAS colours particularly well. The reason for this is that the dust mixture used in Cloudy is not appropriate for carbon-rich planetary nebulae. We have therefore used the final model as input to the Siebenmorgen dust model, which includes stochastically heated small grains and PAHs in a radiative-transfer calculation (Siebenmorgen and Krügel 1992). Both components are expected to give a significant (or even dominant) contribution to the 12$\mu$m continuum flux in carbon-rich planetary nebulae (Siebenmorgen et al. 1993). This model predicts IRAS fluxes in much better agreement with observations, as shown in the Table. However, we note that no total dust mass can be derived unless measurements at much longer wavelengths are available. The shell mass given in the Table should be viewed with caution.

When we compare our results with those from Dopita and Meatheringham (1991) (hereafter DM) significant differences in the shell radius and density are apparent. The main reason is probably our use of the radio flux density rather than the H$\beta$ flux as measure for the recombination rate: clearly the discrepancy between these two values needs to be resolved. Smaller differences can be caused by our exclusion of the HeII $\lambda4686$ line, resulting in a slightly lower stellar temperature. The stellar temperature is in agreement with the spectral type and with the Zanstra temperature. The stellar luminosity agrees very well with DM. The hydrogen density is somewhat higher than determined from the [OII] line ratio (the value DM used), but the line-ratio method shows large uncertainties at high densities. Also the mean electron temperature is a little higher than determined from the [OIII] line ratio. The abundances are in good agreement with the values determined by DM and indicate a non-type I PN (Clegg 1993). The nitrogen abundance seems very low. Also the carbon and neon abundances seem to deviate, but these values are very uncertain. The high (but uncertain) carbon abundance and the agreement of the Siebenmorgen model with the observed IRAS fluxes indicate that the nebula is carbon rich.

The derived inner radius is very small for a planetary nebula, and indicates that the nebula is very young. The ionisation radius implies an angular *diameter* of $0.08''$. Comparing this with the predictions from the radio emission shows that the radio emission is marginally affected by optical-depth effects; the optically thin radio flux would be about 1.0 mJy. This effect is insufficient to explain the discrepancy between the predicted and observed H$\beta$ flux.

We conclude that a very good fit to the observed parameters can be obtained using the radio and IRAS flux densities.

## 6. Comparison with Galactic [WC] stars

Although few [WC] stars are known in the LMC, they are not uncommon among Galactic PN. The IRAS detections allow us to compare SMP58 and SMP61 with their Galactic counterparts. [WC] stars are commonly classified as [WC$n$] where $n$ indicates the degree of excitation of the star: it ranges from $n = 2$ for the hottest ($T_\star \approx 10^5$ K) to $n = 11$ for the coolest stars ($T_\star \approx 2 \times 10^4$ K). SMP58 and SMP61 both are classified as [WC4/5] (Monk et al. 1988), with $T_\star \simeq 60000$ K. The Zanstra temperatures of SMP58 and SMP61 are in reasonable agreement with this, at 67500 K and 48000 K respectively (Meatheringham and Dopita 1991a). The model temperature derived above for SMP58 also agrees well with the excitation temperature.

Fig. 2a shows the distribution of $F_{12}/F_{25}$ flux ratios as a function of subtype, for the two LMC [WC] stars and for all known galactic [WC] stars listed in the ESO–Strasbourg Catalogue of Planetary Nebulae (Acker et al. 1992). We only include reliable (FQ= 3) IRAS detections within $15''$. The colour distribution of the [WC] stars is similar to that of the general PN population (e.g. Pottasch et al. 1988). However, the higher subtypes (cooler stars) show a tendency toward higher ratios. This agrees qualitatively with the expected evolution: during the post-AGB evolution the stars become hotter with time (corresponding to decreasing $n$), while the envelope expands and cools (decreasing $F_{12}/F_{25}$ flux ratio). Thus, an evolutionary sequence can be envisaged connecting the cool to the hot [WC] stars.

In Fig. 2a, the filled circles indicate sources with $F_{12} > 10$ Jy, which roughly corresponds to the IRAS sensitivity for the LMC scaled to the distance of the bulge: all sources represented by the open circles would not be detectable in the LMC. The colours of the two LMC [WC] stars are reasonably similar to those of galactic [WC] stars of similar flux. However, the subtype of the LMC stars is much earlier: $n = 4$–5, where their galactic counterparts have 8–11. In terms of stellar temperature, the two LMC objects are about twice as hot as their Galactic counterparts with similar IRAS colours.

We can assign time scales to the IRAS colours by using the calculated post-AGB evolution for the galactic [WC9] planetary nebula BD+30°3639 (Siebenmorgen et al. 1993). Fig. 2b shows this track in the full colour–colour diagram, where the LMC stars are not plotted as they were not detected at 60$\mu$m. The time scales assume a constant expansion velocity of $22\,\mathrm{km\,s^{-1}}$. The observed range of $F_{12}/F_{25}$ flux ratios for the cool [WC] stars correspond to $t = 200$–1000 yr after the end of the AGB. These short time scales are confirmed by the fact that, of the 15 known [WC8–11] stars, 7 show evidence for dust emission at K (IR-type D, Whitelock 1985). In contrast, of the 21 earlier [WC] stars, none are D-type (using data from Acker et al. 1992 and Zijlstra et al. 1991). K-band dust emission is only expected in objects where the mass loss has ended $\lesssim 300$–500 yr ago (Siebenmorgen et al. 1993). This conclusion takes into account stochastic heating of small grains.

The IRAS flux ratios of both SMP58 and SMP61 also correspond to such young objects, unlike their Galactic counterparts of similar stellar temperature. This indicates that both LMC objects have more compact dust shells than galactic [WC] stars of the same temperature. The model fit for SMP58 discussed above confirms its compactness. Either the stars have evolved quicker, or the nebulae have expanded more slowly than their Galactic counterparts. Support for the latter comes from Wood et al. (1992), who argue that AGB stars in the LMC have lower expansion velocities by about a factor of two than galactic AGB stars, due to the lower metallicity of the LMC. This effect could naturally explain the observed difference between Galactic and LMC [WC] stars.

## 7. Other possible causes for blue IRAS colours

We derive a picture above in which SMP58 contains a very compact nebula which has expanded slowly. However, before making this conclusion too strong, we should warn the reader that, if the blue IRAS colours could be explained in any other way than a compact nebula, the reason to prefer our ionisation model over the more extended nebula derived by Dopita and Meatheringham (1991) would disappear. The compactness of the nebula would then purely depend on whether one accepts the radio flux density or the H$\beta$ flux. It is therefore necessary to mention a few ways in which the low metallicity of the LMC could cause bluer IRAS colours, other than causing lower expansion velocities.

First, line blanketing in the far-UV could be lower for the LMC central stars. The importance of this is very difficult to estimate, since the atmosphere of [WC] stars is highly self-enriched in some elements (arising from the s-process) but not in others. However, if the line blanketing is significantly lower it would cause both the Zanstra temperature and the [WC] excitation temperature to be somewhat overestimated, which would reduce the discrepancy in IRAS colours between the LMC and Galactic [WC] stars. This points requires detailed calculations which are not yet available.

Second, the IRAS bands contain contributions both from dust continuum and a number of forbidden lines. If the dust content is lower, the lines could become more dominant; this would make the 12$\mu$m flux relatively stronger since it contains more emission lines, where especially [NeII] 12.81$\mu$m can be strong. However, the emission lines are significant for the total flux only in old, evolved nebulae. For SMP58, Cloudy gives an emission line contribution of 0.16% in the 12$\mu$m band and 0.02% in the 25$\mu$m band.

Finally, the dust composition could be significantly different from the composition in Galactic PN. To explain the blue IRAS colours, one would need a larger contribution of the smallest grains, especially PAHs. It is not clear why a lower dust content would lead to smaller grains—the grain composition is probably largely determined by destruction mechanisms, both for Galactic and LMC PN.

Of these four effects, a lower expansion velocity seems to give the more likely explanation for the observed blue IRAS colours. However, the other effects deserve further investigation.

## 8. Conclusions and future prospects

We have presented the first radio continuum detection of a PN in the LMC. The radio flux density is considerably lower than expected from the observed H$\beta$ flux. The object, which contains a [WC] star, was also detected by IRAS. We show that all IRAS-detected PN, are likely to be young objects. We also compare the PN to [WC]-type objects in the galaxy, and find that the LMC PN has much bluer IRAS colours than its Galactic counterparts. Our preferred explanation is that its circumstellar envelope is more compact due to a lower expansion velocity, caused by lower metallicity. This is supported by a photo-ionisation model fitting of the spectrum of SMP58 However, other possible causes also deserve further study.

Our radio continuum detection indicates that many PN in the LMC could be detected in the radio, since SMP58 is not particularly luminous. This raises the possibility of obtaining accurate stellar luminosities for luminous, high-mass PN. Such a study has not been possible in our Galaxy because of the lack of such objects with known distances; as a consequence, at present it is not known what fraction of PN have massive central stars (e.g. Pottasch 1992, Dopita et al. 1993). This discussion also has implications for the mass distribution of white dwarfs, which originate from PN central stars. A radio survey of LMC PN appears the best way to resolve this controversy.

**Acknowledgements**

We thank the staff at the Australia telescope, especially Wilfred Walsh and Neil Killeen, for their support and advice during the observations and data reduction. We also thank Helen Pongracic of Sydney University for assisting with the data reduction. Rens Waters is acknowledged for several suggestions regarding the effects of low metallicity. Rene Oudmaijer helped us with accessing the IRAS data base. Tim Bedding, Griet van de Steene and Stuart Pottasch are thanked for critical reading of the manuscript. Peter van Hoof receives financial support under grant nr. 782-372-033 from the Netherlands Foundation for Research in Astronomy (ASTRON). This research has made use of the SIMBAD database, operated by CDS, Strasbourg, France, and of the Cloudy program, obtained from the University of Kentucky, USA.

**Figure captions**

**Figure 1.** The 3cm radio image of SMP58, from the June 1993 observations (Table 1). Contour levels are at $-0.4, -0.2, 0.2, 0.4, 0.6, 0.8$ mJy beam$^{-1}$.

**Figure 2.** a. Excitation class versus IRAS 12/25$\mu$m flux ratio for [WC] stars. Open circles: galactic stars with $F_{12} < 10$ Jy; filled circles: galactic stars with $F_{12} > 10$ Jy; triangles: LMC [WC] stars. b. IRAS colour evolution of the galactic [WC9] star